\documentclass[aps,graphicx,twocolumn,groupedaddress,superscriptaddress]{revtex4-2}%
\usepackage{amsmath}
\usepackage{graphicx}
\usepackage{url}
\usepackage[colorlinks]{hyperref}
\usepackage{CJK}
\usepackage{graphics,graphicx,float,amsmath, xcolor} 
\hypersetup{%
    plainpages=true,
    breaklinks=true,%not default in dvips mode, so we must specify
    hypertexnames=false,%not ideal, but needed when pagenums duplicate (`i' vs. `1')
    pageanchor=true,
    colorlinks=true,
    linkcolor={blue},
    citecolor={blue},
    urlcolor={blue},
    anchorcolor={black}
}

\newcommand{\beq}{\begin{equation}}
\newcommand{\eeq}{\end{equation}}

\allowdisplaybreaks[1]
\begin{document}

%\preprint{}

\title{Measurement-device-independent quantum-secret-sharing networks with linear Bell-state analysis}

\author{Tianqi Liu}
\affiliation{MIIT Key Laboratory of Semiconductor Microstructure and Quantum sensing, School of Physics,  Nanjing University of Science and Technology, Nanjing {\rm 210094}, China}
\author{Jiancheng Lai}
\affiliation{MIIT Key Laboratory of Semiconductor Microstructure and Quantum sensing, School of Physics,  Nanjing University of Science and Technology, Nanjing {\rm 210094}, 
China}
\affiliation{Engineering Research Center of Semiconductor Device Optoelectronic Hybrid Integration in Jiangsu Province, Nanjing {\rm 210094}, China}
\author{Zhenhua Li}
\affiliation{MIIT Key Laboratory of Semiconductor Microstructure and Quantum sensing, School of Physics,  Nanjing University of Science and Technology, Nanjing {\rm 210094}, 
China}
\affiliation{Engineering Research Center of Semiconductor Device Optoelectronic Hybrid Integration in Jiangsu Province, Nanjing {\rm 210094}, China}
\author{Tao Li}
\email[]{tao.li@njust.edu.cn}
\affiliation{MIIT Key Laboratory of Semiconductor Microstructure and Quantum sensing, School of Physics,  Nanjing University of Science and Technology, Nanjing {\rm 210094}, China}
\affiliation{Engineering Research Center of Semiconductor Device Optoelectronic Hybrid Integration in Jiangsu Province, Nanjing {\rm 210094}, China}
\date{\today}

%\footnote{$\dagger$ These authors contributed equally to this work.}

\begin{abstract}
Quantum secret sharing (QSS) plays a pivotal role in multiparty quantum communication, ensuring the secure distribution of private information among multiple parties. However, the security of QSS schemes can be compromised by attacks exploiting imperfections in measurement devices. Here, we propose a reconfigurable approach to implement QSS based on measurement-device-independent (MDI) principles, utilizing linear two-photon Bell state analysis.
By employing single-qubit conjugate operations for encoding private classical information, our approach offers reconfigurability, allowing for the inclusion of additional parties without sacrificing efficiency. Furthermore, we demonstrate the robust security of our MDI-QSS scheme against inter-eavesdropping by dishonest participants and establish lower bounds for secure communication among three legitimate parties. This work presents a flexible configuration for implementing multiparty secure quantum communication with imperfect measurement devices and represents a significant advancement in the development of secure quantum communication technologies.
\end{abstract}
\maketitle
\date{\today}

\section{Introduction}

Quantum communication utilizes quantum physical principles to provide unconditional security for communication parties~\cite{gisin2007quantum}. Quantum key distribution~(QKD)~\cite{lo2014secure,Santo2018Two-Way,Shang2020oneway,Xie2022Breaking,Zhou2024Sending-or-not-sending} and quantum secure direct communication~(QSDC)~\cite{long2002theoretically,deng2003two, Shapiro2019,ZHOU202012,Li2020QSDC,Sheng2022One-step} are two prominent frameworks that enable point-to-point quantum communication and have demonstrated remarkable performance in practical applications~\cite{Chen2021integrated,Wang2022Twin-field,Li2023High-rate,hu2016experimental,zhang2017quantum,Massa2019Experimental, Qi202115user}. In contrast, quantum secret sharing (QSS) is a multiparty quantum communication protocol~\cite{Hillery1999,Karlsson1999,Xiao2004QSS, Zhang2005QSS,Zhang2005Multiparty,ting2005controlled,gao2005deterministic,Qin2007Cryptanalysis} that enables the secure distribution of a secret among multiple parties, allowing only a designated subset of these parties, when collaborating, to reconstruct the secret~\cite{ Kogias2017Unconditional, Brian2019Quantum,Habibidavijani_2019,Gu2021Differential,wu2020Passive,  Ouyang2023Approximate, Qin2024Efficient,Tian2024Dynamic, Zhang2024Device-independent, Xiao2024Source-independent,Zhang2025Device-independent}. 
In principle, various QSS protocols can be categorized into the following two main branches: (1) sharing classical information among designated parties through quantum channels~\cite{Hillery1999,Karlsson1999,Xiao2004QSS, Zhang2005QSS,Zhang2005Multiparty,Zhang2023Multiple-participant,ting2005controlled,gao2005deterministic,Qin2007Cryptanalysis,Kogias2017Unconditional, Habibidavijani_2019,Brian2019Quantum,Gu2021Differential,wu2020Passive,  Ouyang2023Approximate, Qin2024Efficient,Tian2024Dynamic, Zhang2024Device-independent, Xiao2024Source-independent,Zhang2025Device-independent}; (2) sharing an unknown quantum state among parties, where no information can be revealed with fewer than the designated number of parties~\cite{Hillery1999,Cleve1999How,LI2004Multiparty,Lu2016Secret,Modi2018Masking,Liang2020Impossibility,Singh2024Controlled}, also known as quantum state sharing~\cite{Lance2005Continuous-variable,Deng2005Multiparty,Wilkinson2023Quantum}. 
However, the practical implementation of quantum communication protocols is often vulnerable to side-channel attacks due to imperfect apparatuses~\cite{Xu2020Secure}, particularly in scenarios where the trustworthiness of measurement devices cannot always be guaranteed~\cite{Chen2005Experimental,Schmid2005Experimental, Gaertner2007Experimental,Zhou2018Quantum,Shen2023Experimental}. These vulnerabilities represent a significant threat to the security of real-world quantum communication.

%Lu2016Secret, 

Measurement-device-independent~~(MDI) quantum communication~\cite{Xu2020Secure} addresses this challenge and closes all potential loopholes for side-channel attacks by using postselected entanglement. It was initially introduced to implement MDI QKD protocols~\cite{lo2012measurement,Braunstein2012Side-channel,
Ma2012Alternative, Xu2013Practical,Sun2023Practical}  and then extended to perform MDI-QSDC protocols~\cite{Li2020MDIreview,zhou2020measurement, gao2019long,Ying2022Measurement-device-independent} for directly transmitting private information over quantum channels. MDI-QSS protocols~\cite{Fu2015Long-Distance,Gao2020Deterministic, Das2021Universal, Wei2022Sender-controlled,Ju2022Measurement-device-independent,Li2023Breaking, zhang2024memory}  ensure the security of the secret sharing process among multiple parties by postselecting multiphoton Greenberger–Horne–Zeilinger~(GHZ) states~\cite{Liu2022Universal,Zhou2023Parallel,Li2024Heralded} and their security is independent of the results of the measurement devices involved. The essential building 
block of MDI quantum communication is entangled state analysis~\cite{Sheng2010Complete,Ren2013Hyperentanglement, Bhaskar2020Experimental}, which consists of linear optical elements and single-photon detectors with an upper bound probability of $1/2^{(k-1)}$ for implementing a $k$-photon GHZ-state analyzer~(GSA)~\cite{Pan1998GHZanalyzer,Lu2009swap,Guus2023Analysis}. The probability can, in principle,  be improved to unity~\cite{Li2019Resource-efficient} when complex deterministic interfaces between single photons and individual atoms are used~\cite{reiserer2015cavity,li2018gate,Beukers2024Remote-Entanglement}.

Here, we present a scalable and secure MDI-QSS protocol that uses a two-photon Bell-state analyzer~(BSA)~\cite{Sheng2010Complete,Ren2013Hyperentanglement,Bhaskar2020Experimental} rather than multiphoton GSAs~\cite{Pan1998GHZanalyzer,Lu2009swap,Guus2023Analysis}, which involve the interference of multiple photons generated by different sources~\cite{Ruf2021Quantum,Zhang2021Spontaneous,Wang2022Self-healing}. Our protocol uses an entangled photon pair  as a data bus. Each communication party randomly encodes their private classical information by performing one of two pairs of conjugate unitary operations on the photon  arriving at their respective node. The photon pair is then measured by a BSA at an untrusted ancillary node. Based on the BSA results, the legitimate parties can share the required correlations for implementing QSS. A subset of the parties can collaborate to determine the private classical secret of the remaining party using practical measurement devices. Furthermore, we demonstrate the security of our MDI-QSS protocol by using the Holevo theorem~\cite{holevo1973bounds} and present analytical QSS generation rates for implementing a three-party MDI-QSS protocol involving a dishonest participant~\cite{Qin2007Cryptanalysis,Kogias2017Unconditional,Brian2019Quantum}. 
Our MDI-QSS protocol provides a flexible configuration for implementing multiparty quantum communication and represents a significant advancement in the development of robust quantum communication technologies.

\section{Protocol description}

\begin{figure}[t!]
\includegraphics[width=0.48\textwidth]{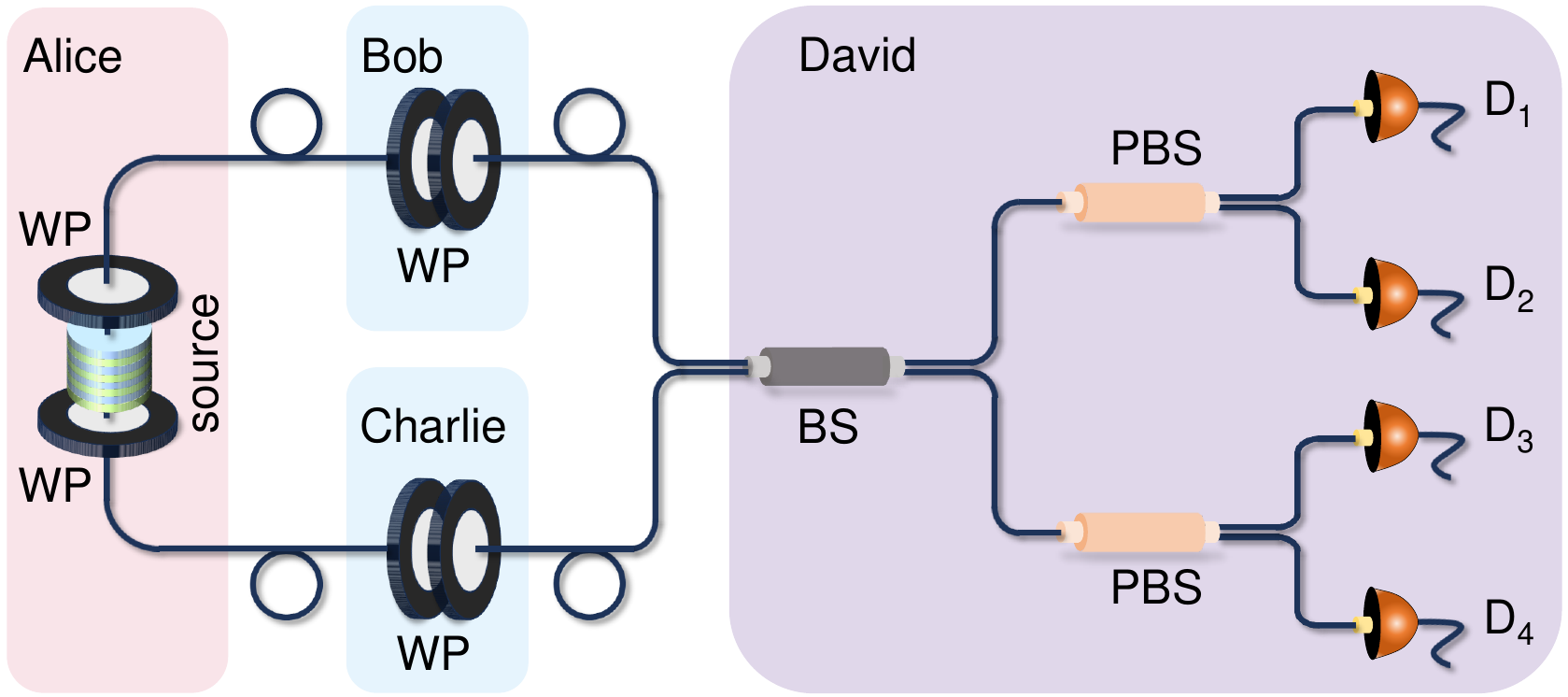}
\caption{
Schematic of a three-party MDI-QSS protocol. WP denotes a waveplate unite. BS denotes a balanced beam splitter, while PBS denotes a polarizing beam splitter that transmits~(reflects) photons in state $\left| H \right\rangle$~($\left| V \right\rangle$). ${D_i}$ ($i = 1,2,3,4$) denotes a practical single-photon detector.
}
\label{Fig1}
\end{figure}
\subsection{Three-party MDI-QSS protocol} \label{secIIA}
~The schematic of our MDI-QSS protocol for three parties is shown in Fig.~\ref{Fig1} and consists of four individual nodes. Alice, Bob, and Charlie can share private information with the help of an untrusted node, David. Alice randomly prepares one photon pair in one of four Bell states in two conjugate bases.  She then sends one photon to Bob and the other photon to Charlie. Upon receiving the photons, Bob and Charlie each perform one of four unitary operations, exchanging the state of the photon pair among the four Bell states, and send the photons to David for Bell-state measurements using a BSA~\cite{Xu2020Secure}.  In practice, a single-photon filter is implicitly incorporated prior to the operations~\cite{Li2020QSDC}. The BSA can be constructed using linear optical elements and single-photon detectors and distinguish two Bell states~\cite{lo2012measurement,Braunstein2012Side-channel,
Ma2012Alternative, Xu2013Practical}. Based on the result of the BSA, Alice, Bob, and Charlie can share private information by postselecting cases that lead to a deterministic result of the BSA. Specifically, the MDI-QSS protocol is carried out in the following steps.

(1)~Alice prepares an entangled photon pair in the Bell state $\left| {{\psi^{+}}} \right\rangle  = (\left| {HV} \right\rangle  + \left| {VH} \right\rangle )/\sqrt{2}$ and then randomly evolves them into one of four Bell states, across two conjugate bases with
\begin{equation}
	\begin{aligned}
 	\left| {{\psi^{\pm}}} \right\rangle  &= \frac{1}{{\sqrt 2 }}(\left| {HV} \right\rangle  \pm \left| {VH} \right\rangle ),\\
 	\left| {{\varphi^{\pm}}} \right\rangle  &= \frac{1}{{\sqrt 2 }}(\left| {HV} \right\rangle  \pm i \left| {VH} \right\rangle ).\\
 	\label{eq1}
 	\end{aligned}
 \end{equation}
This can be achieved by passing the photons through wave-plate~(WP) units that are tuned to introduce one of the following local operations: 
\begin{eqnarray}
	{U_{00}} &=& \left| H \right\rangle \left\langle H \right| + \left| V \right\rangle \left\langle V \right|,\nonumber \\
	{U_{01}} &=& \left| H \right\rangle \left\langle H \right| - \left| V \right\rangle \left\langle V \right|,\nonumber\\
	{U_{10}} &=& \left| H \right\rangle \left\langle H \right| + i\left| V \right\rangle \left\langle V \right|,\nonumber \\
	{U_{11}} &=& \left| H \right\rangle \left\langle H \right| - i\left| V \right\rangle \left\langle V \right|. \label{eqUnitary}
\end{eqnarray}
The operator ${U_{ab}}$ exchanges the states within the same basis, $\left| {{\psi^{\pm}}} \right\rangle\leftrightarrow\left| {{\psi^{\mp}}} \right\rangle$ and $\left| {{\varphi^{\pm}}} \right\rangle\leftrightarrow\left| {{\varphi^{\mp}}} \right\rangle$, for $a=0$, and changes the states into the other basis, $\left| {{\psi^{\pm}}} \right\rangle\leftrightarrow\left| {{\varphi^{\pm}}} \right\rangle$ and $\left| {{\varphi^{\pm} }} \right\rangle\leftrightarrow\left| {{\psi^{\pm}}} \right\rangle$, for $a=1$, while the superscript flips for $b=1$ and remains unchanged for $b=0$. Subsequently, Alice sends one photon to Bob and sends the other photon to Charlie.

(2)  Bob (Charlie) randomly encodes the private information on the photon arriving at his node by directing it into a WP unit, which implements one of the four local operations ${U_{a_ib_i}}$, as shown in Eq.~(\ref{eqUnitary}). For simplicity, we use a pair of binary  $\{ {a_i},{b_i}\}$, where the subscript $i=A,~B,$ and $C$ distinguishes the operations performed by Alice, Bob, and Charlie, respectively. Therefore, $U_{a_ib_i}$ for ${a_i}=1$ transforms the state from one basis into another, while $U_{a_ib_i}$ for $b_i=1$ modifies the superscripts, as illustrated at Alice's node. After both parties have completed encoding, they send their photons to the untrusted party David  for measuring  in the Bell-state basis.

(3) David measures the received photon pair using a BSA composed of linear optical elements and single-photon detectors~\cite{lo2012measurement,Braunstein2012Side-channel,
Ma2012Alternative, Xu2013Practical}. The BSA is  configured  to deterministically distinguish~$\left| {{\psi ^ + }} \right\rangle $ and $\left| {{\psi ^ - }} \right\rangle $, while producing unbiased outcomes for a photon pair in Bell states $\left| {{\varphi^\pm }} \right\rangle$, where
\begin{eqnarray}
\left| {{\varphi ^ + }} \right\rangle &=& \frac{{1 + i}}{2}\left| {{\psi ^ + }} \right\rangle  + \frac{{1 - i}}{2}\left| {{\psi ^ - }} \right\rangle , \nonumber\\
	\left| {{\varphi ^ - }} \right\rangle  &=& \frac{{1 - i}}{2}\left| {{\psi ^ + }} \right\rangle  + \frac{{1 + i}}{2}\left| {{\psi ^ - }} \right\rangle.
\end{eqnarray}

(4) The first three steps are repeated many times until David gets $n$ results of the BSA. Subsequently, David publicly discloses the results of the BSA.

(5) After receiving the measurement results, Alice randomly selects ${n_s}$ photon pairs as samples and publicly discloses their positions. Subsequently,  Bob and Charlie randomly disclose the corresponding $\{ {a_B},{b_B}\}$ and  $\{ {a_C},{b_C}\}$  in any order they choose. Alice discards all data ${a_A} \oplus {a_B} \oplus {a_C} = 1$, and then analyzes the error rate of the remaining data to detect potential eavesdropping in the MDI-QSS protocol. If the error rate exceeds a predefined threshold, communication is immediately terminated. Otherwise, the protocol proceeds to the next process. In the ideal scenario, the correlations of the entangled states prepared by Alice, the encoding operations performed by Bob and Charlie, and the BSA results disclosed by David are shown in Tables~\ref{table1} and~\ref{table2}. The BSA results are deterministic for cases where the public keys $a_i$ satisfy the condition ${a_A} \oplus {a_B} \oplus {a_C} = 0$. These $a_i$ are then used for security analysis and key generation based on the correlation between the private keys $b_i$ and the BSA results. Specifically, we have $s={a_A} \oplus{b_A} \oplus {b_B} \oplus {b_C}$, where $s=0$~($s=1$) refers to the BSA result $|\psi^+\rangle$ ($|\psi^-\rangle$). For cases where ${a_A} \oplus {a_B} \oplus {a_C} = 1$, the BSA results $s=0$ and $s=1$ occur with   equal probability. These results are denoted by ``$ - $" and should be discarded in subsequent analyses.

\begin{table}[t!]
	\centering
\renewcommand{\arraystretch}{1.5}
	\caption{BSA results versus  operations performed by Bob and Charlie when Alice prepares Bell states $\left| {{\psi ^ \pm }} \right\rangle $ with $U_{0_A0_A}$ and $U_{0_A1_A}$.}
 \setlength{\tabcolsep}{3mm}{
 \begin{tabular}{l cccc}
	\hline
\hline
&${U_{0_B0_B}}$&${U_{0_B1_B}}$&${U_{1_B0_B}}$&${U_{1_B1_B}}$\\
	 \hline

	${U_{0_C0_C}}$ &$\left| {{\psi ^ \pm }} \right\rangle $&$\left| {{\psi ^ \mp }} \right\rangle $&$ -   $&$ -  $ \\
    ${U_{0_C1_C}}$ &$\left| {{\psi ^ \mp }} \right\rangle $&$\left| {{\psi ^ \pm }} \right\rangle $&$ -   $&$ -   $ \\
	${U_{1_C0_C}}$ &$ -   $&$ -   $&$\left| {{\psi ^ \pm }} \right\rangle $&$\left| {{\psi ^ \mp }} \right\rangle $\\	
	${U_{1_C1_C}}$ &$ -   $&$ -   $&$\left| {{\psi ^ \mp }} \right\rangle $&$\left| {{\psi ^ \pm }} \right\rangle $ \\
 \hline
 \hline
\end{tabular}\label{table1}
}
\end{table}
\begin{table}[t!]
	\centering
\renewcommand{\arraystretch}{1.5}
	\caption{BSA results versus  operations performed by Bob and Charlie when Alice prepares Bell states $\left| {{\varphi ^ \pm }} \right\rangle $ with $U_{1_A0_A}$ and $U_{1_A1_A}$.}\label{table2}
 \setlength{\tabcolsep}{3mm}{	\begin{tabular}{l cccc}		
		\hline
\hline
		&${U_{0_B0_B}}$&${U_{0_B1_B}}$&${U_{1_B0_B}}$&${U_{1_B1_B}}$\\
		\hline
		${U_{0_C0_C}}$ &$ -   $&$ -   $&$\left| {{\psi ^ \mp }} \right\rangle $&$\left| {{\psi ^ \pm }} \right\rangle $ \\
		${U_{0_C1_C}}$ &$ -   $&$ -   $&$\left| {{\psi ^ \pm }} \right\rangle $&$\left| {{\psi ^ \mp }} \right\rangle $\\	
		${U_{1_C0_C}}$ &$\left| {{\psi ^ \pm }} \right\rangle $&$\left| {{\psi ^ \mp }} \right\rangle $&$ -   $&$ -   $ \\
		${U_{1_C1_C}}$ &$\left| {{\psi ^ \mp }} \right\rangle $&$\left| {{\psi ^ \pm }} \right\rangle $&$ -   $&$ -   $ \\
		\hline
\hline
	\end{tabular}}
\end{table}

(6)~ After the security check, Alice, Bob, and Charlie publicly disclose their public keys~(${a_A}$, ${a_B}$, ${a_C}$) of the remaining $n - {n_s}$ photons in their hands and discard all cases that do not satisfy the condition ${a_A} \oplus {a_B} \oplus {a_C} = 0$. At this stage, the three parties will generate a key string with an approximate length of ${{(n - {n_s})} \mathord{\left/
		{\vphantom {{(n - {n_s})} 2}} \right.
		\kern-\nulldelimiterspace} 2}$, where each key satisfies the correlation ${b_A} \oplus {b_B} \oplus {b_C}={a_A} \oplus s$. Two parties~(e.g., Bob and Charlie) can collaborate to determine the private key of the remaining party~(e.g., Alice).  Subsequently, postprocessing techniques, such as error correction and privacy amplification~\cite{Xu2020Secure}, are applied to obtain the final key.

\subsection{Four-party MDI-QSS protocol} \label{secIIB}
In principle, the three-party MDI-QSS protocol can be generalized to share private classical information among more than three parties. For implementing a four-party MDI-QSS protocol with a two-photon BSA, the process can be described as follows:\\
\indent
(1) Alice  prepares one photon pair randomly in one of four Bell states $| {{\psi ^ \pm }}\rangle$ and $| {{\varphi ^ \pm }}\rangle$. She then sends one photon to Bob and the other photon to Charlie. \\
\indent
(2) Upon receiving these photons, Bob and Charlie each encode their private information by performing one of four unitary operations, based on their randomly chosen public and private keys $\{a_i, b_i\}$. Bob sends his photon to the untrusted party David, while Charlie sends his photon to the fourth communication party, Daniel.\\
\indent 
(3) Daniel then encodes his private information by performing one of four unitary operations, as described in Eq.~(\ref{eqUnitary}), on the photon he received, based on his own randomly chosen public and private keys $\{a_D, b_D\}$.  Daniel then sends the photon to the untrusted party David, who performs a BSA measurement on the two photons arriving at his node.\\
\indent
(4) After repeating the first three steps many times, David publicly declares the results of all BSA measurements.\\
\indent
(5) Alice randomly selects some photon pairs as samples and publicly discloses their positions. Subsequently, she asks Bob, Charlie, and Daniel to randomly disclose their  corresponding public and private keys $\{a_B, b_B\}$, $\{a_C, b_C\}$, and $\{a_D, b_D\}$. She discards the cases with  ${a_A} \oplus {a_B} \oplus {a_C} \oplus {a_D} = 1$, and analyzes the error rate of the remaining cases with  ${a_A} \oplus {a_B} \oplus {a_C} \oplus {a_D} = 0$. The correlations established between the private keys of four parties, after postselecting the BSA results, are ${a_A} \oplus{b_A} \oplus {b_B} \oplus {b_C} \oplus {b_D} = s$, where  $s=0$~($s=1$) refers to the BSA result $| {{\psi^+}}\rangle$~($| {{\psi^-}}\rangle$). \\
\indent
(6) After ascertaining the security, Alice, Bob, Charlie, and Daniel publicly disclose their public keys 
of the remaining photons in their hands and discard all cases that do not satisfy the condition  ${a_A} \oplus {a_B} \oplus {a_C} \oplus {a_D} = 0$. At this stage, the four parties will generate a key string where each key satisfies the correlation ${b_A} \oplus {b_B} \oplus {b_C} \oplus {b_D} = s \oplus{a_A}$. Specifically, any three parties can collaborate to determine the private key of the fourth party. 

\section{Security analysis}

In this section, we present the security analysis of our three-party MDI-QSS protocol. We provide the upper bound on the amount of information that can be obtained by an eavesdropper using collective attacks and present the QSS generation rate for practical parameters. In collective attacks, the eavesdropper, Eve, attaches one probe per qubit and can  measure several of these probes coherently~\cite{Pirandola2020Advances}. The accessible information available to Eve is upper-bounded by her Holevo information on the target variable~\cite{Pirandola2020Advances}. In principle, collective attacks are shown to be as efficient as coherent attacks on quantum communication for the permutation symmetry of the classical postprocessing~\cite{Renner2009de}. Furthermore, a reduction from general attacks to collective attacks is possible by using phase-space symmetries and the postselection technique~\cite{Leverrier2013Security}.

We consider a potentially stronger internal eavesdropping scenario~\cite{Qin2007Cryptanalysis,Kogias2017Unconditional,Brian2019Quantum}, where one party, Bob, is assumed to be dishonest. Bob attempts to obtain private information by eavesdropping on the quantum channel, in collaboration with Eve. We show that the collaboration between Bob and Eve can not determine the private information of Alice. For simplicity, we disregard Bob's actions and assume that he only implements the operation $U_{00}$, since he is dishonest and always shares his operation information $U_{a_Bb_B}$~(i.e., public and private keys $\{a_B, b_B\}$) with Eve. In this scenario, the security of the three-party QSS is simplified to the security of the QKD between the two honest parties~(Alice and Charlie), in the presence of a dishonest party~(Bob)~\cite{Kogias2017Unconditional,Brian2019Quantum}.

In our three-party MDI-QSS protocol, Alice randomly prepares each photon pair in one of the four Bell states, as shown in Eq.~(\ref{eq1}), with an equal probability as the data bus to connect all nodes for secret sharing. Consequently, each photon pair prepared by Alice can be regarded as being in a mixed state~\cite{qi2019implementation} 
\begin{equation}
		\begin{aligned}
	\rho  &= \frac{1}{2}(\left| {{\psi ^ + }} \right\rangle \left\langle {{\psi ^ + }} \right| + \left| {{\psi ^ - }} \right\rangle \left\langle {{\psi ^ - }} \right|)
	\\
	&= \frac{1}{2}(\left| {{\varphi ^ + }} \right\rangle \left\langle {{\varphi ^ + }} \right| + \left| {{\varphi ^ - }} \right\rangle \left\langle {{\varphi ^ - }} \right|),
	\label{eq4}
		\end{aligned}
\end{equation}
with the two photons sent to Bob and Charlie, respectively.

Considering a collective attack scenario, where Bob is a dishonest participant and always informs Eve of his unitary operation, Eve's focus is on eavesdropping on the forward channel from Alice to Charlie. The most general quantum operation that Eve can perform on this forward channel involves a joint operation ${U_C}$ on the photon and an ancilla $\left| \varepsilon  \right\rangle $ belonging to Eve~\cite{qi2019implementation,Xu2020Secure}. The combined system, consisting of two photons and the ancilla, evolves into
\begin{equation}
{\rho ^{AE}} = {U_C}(\rho  \otimes \left| \varepsilon  \right\rangle \left\langle \varepsilon  \right|)U_C^\dag.
\end{equation}
Then, Eve forwards the photon to Charlie while keeping the ancilla. After receiving the photon, Charlie randomly encodes it with one of four $U_{a_Cb_C}$ operations, each with equal probability $p=0.25$. 

After both parties have completed their encoding operations, the combined system consisting of the photon pair and the ancilla evolves into \begin{equation}
{\rho ^{ACE}} = \frac{1}{4}(\rho _{00}^{AE} + \rho _{01}^{AE} + \rho _{10}^{AE} + \rho _{11}^{AE}), \label{rhoACE}
\end{equation}
where we have omitted the identity operation $U_{0_B0_B}$ and the sub-subscript of $\rho _{ab}^{AE} = {U_{a_Cb_C}}{U_C}(\rho  \otimes \left| \varepsilon  \right\rangle \left\langle \varepsilon  \right|)U_C^\dag U_{a_Cb_C}^\dag$, for simplicity of notation. According to Tables~\ref{table1} and ~\ref{table2},
only cases where ${a_A}\oplus{a_C} = 0$ and ${a_B}=0$ contribute to the key generation. When Charlie applies the operation ${U_{00}}$ or ${U_{01}}$,  the photon-pair state is considered to be in the $X$ basis, $\rho  = (\left| {{\psi ^ + }} \right\rangle \left\langle {{\psi ^ + }} \right| + \left| {{\psi ^ - }} \right\rangle \left\langle {{\psi ^ - }} \right|)/2$. Conversely, when  Charlie applies the operation ${U_{10}}$ or ${U_{11}}$,   the photon pair is considered to be in the $Y$ basis, $\rho  = (\left| {{\varphi ^ + }} \right\rangle \left\langle {{\varphi ^ + }} \right| + \left| {{\varphi ^ - }} \right\rangle \left\langle {{\varphi ^ - }} \right|)/2$. 

The upper bound on the mutual information between Alice and Eve can be obtained by the Holevo theorem~\cite{holevo1973bounds} and is described as
	\begin{align}
	I(A:E)& \le \mathop {\max }\limits_{\{ U\} } \{ S({\rho ^{ACE}}) - \frac{1}{4}S(\rho _{00}^{AE})\nonumber \\
	 &~~~- \frac{1}{4}S(\rho _{01}^{AE}) 	- \frac{1}{4}S(\rho _{10}^{AE}) - \frac{1}{4}S(\rho _{11}^{AE})\},
\end{align}
where $S(\rho)=-{\rm Tr}(\rho{\rm log}_2\rho)$ 
denotes the von Neumann entropy. Furthermore, the maximum mutual information between Alice and Eve, as shown in the~\ref{appA}, can be specified as
\begin{equation}
I(A:E) \le \mathop {Q_{\rm Eve}} h({e_x} + {e_y}),
\end{equation}
where $h(x)=-x {\rm log}_2x-(1-x){\rm log}_2(1-x)$ is the binary Shannon entropy; $e_x$~($e_y$) is the bit-flip error in the $X$~($Y$) basis, corresponding to the probability that a photon pair, prepared by Alice in the Bell state $\left| {{\psi ^ \pm }} \right\rangle$~($\left| {{\varphi ^ \pm }} \right\rangle $), flips to the Bell state $\left| {{\psi ^ \mp }} \right\rangle$~($\left| {{\varphi ^ \mp }} \right\rangle $) when reaching Charlie's node, $Q_{\rm Eve}$ represents the maximum  probability that Eve can access the photon pairs.

For an error rate $e$ occurring during photon-pair transmission and measurement processes, thereby violating the correlations outlined in Tables~\ref{table1} and~\ref{table2}, the final QSS generation rate for three parties can  be described as 
\begin{align}
		R &= I(A:BC) - I(A:E) \nonumber\\
		&= Q[1 - h(e)] - {Q_{\rm Eve}}h({e_x} + {e_y}), \label{keyRate}
	\end{align}
where $Q$ represents the probability that the photon pairs produced by Alice reach David's node and are successfully click the single-photon detectors there. In the ideal scenario, $Q = {Q_{\rm Eve}} = 1$ can be achieved under the assumption of perfect quantum channels and flawless single-photon detectors. However, in practical applications,  inherent imperfections, such as photon loss during channel transmission and the nonunity efficiency of single-photon detectors, must be considered.

The success of the BSA applied on two photons in node David is heralded by the simultaneous click of two single-photon detectors. The loss of either photon  leads to an erroneous response of the detectors, thereby preventing the key generation among three communication parties. The efficiency of generating a $1$-bit key with the transmission of a single photon pair is 
\begin{eqnarray}
\eta_0= {\eta^4 _t}{\eta^2 _d}\eta _D,
\end{eqnarray}
where ${\eta _d}$ denotes the efficiency of the single-photon detectors, $\eta_D$ denotes the internal efficiency of David's apparatus, and ${\eta _t}$ denotes the photon transmission efficiency, i.e., ${\eta _t} = {10^{ - {{\alpha L} \mathord{\left/{\vphantom {{\alpha L} {10}}} \right.
\kern-\nulldelimiterspace} {10}}}}$. Here $\alpha $ denotes the channel loss rate,  and  $L$ is the channel length,  assumed to be equal for the four quantum channels. Therefore, we can take $Q = {Q_{\rm Eve}} = \eta_0$ for practical analysis. 

 \begin{figure}[t!]
	\includegraphics[width=0.48\textwidth]{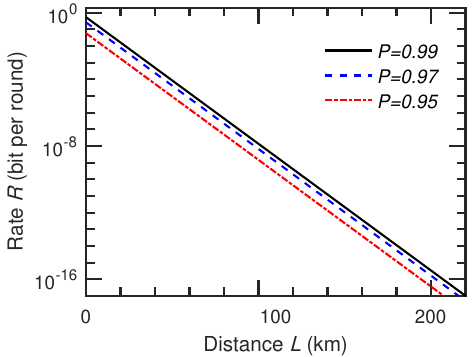}
	\caption{QSS generation rate $R$ versus the communication distance $L$.  Here, we set ${\eta _D} = 86.3\%$, ${\eta _d} = 93\%$, $\alpha= 0.19~{\rm{dB/km}}$, and $P=(1-e)$ is the channel fidelity.}
	\label{Fig2}
\end{figure}

The relationship between the error rates $e$, $e_x$, and $e_y$ can be  expressed as follows:
\begin{align}
	e &= {e_x}(1 - {e_x}) + \frac{1}{2}{e_y}(1 - {e_x}) + \frac{1}{2}(1 - {e_y}){e_x}\nonumber\\
	&= \frac{3}{2}{e_x} + \frac{1}{2}{e_y} - e_x^2 - {e_x}{e_y},\label{general-e}
	\end{align}
where an error in the  $X$-basis or  $Y$-basis during transmission from Alice to Bob and Charlie, followed by an $X$-basis error, leads to a correct BSA result, while an error in the $X$-basis or $Y$-basis contributes to $e$. 
For collective symmetric attacks~\cite{Pirandola2020Advances}, we consider unbiased error rates ${e_x} = {e_y}$, which also take place in dephasing channels in the $Z$ basis~\cite{Xu2020Secure}. The overall system error rate  simplifies to
\begin{equation}
e = 2{e_x}(1 - {e_x}). \label{eqe}
\end{equation}
We note that the following analysis can be generalized to the cases with asymmetric error rates ${e_x} \neq {e_y}$ and the corresponding QSS generation rate is then determined by combining 
Eqs.~(\ref{keyRate}) and (\ref{general-e}).  
The  QSS generation rate of our MDI-QSS protocol, after inserting Eq.~(\ref{eqe}), can be expressed as  
\begin{equation}
R = Q[1 - h(e) - h(1 - \sqrt {1 - 2e} )], \label{Rqss}
\end{equation}
which, as a function of the communication distance $L$ for three different error rates, is shown in Fig.~\ref{Fig2}. Here, we assume that the internal efficiency of David's apparatus is ${\eta _D} = 86.3\%$ (i.e., an internal loss of 0.64dB)~\cite{Li2023High-rate}, the efficiency of the detectors is ${\eta _d} = 93\% $~\cite{Marsili2013Detecting,Li2023Breaking},  and the loss rate of the fiber is $\alpha  = 0.19~\rm{ dB/km}$~\cite{Li2023High-rate} for a telecommunication wavelength of
around $1550$~nm. The QSS generation rate $R$ decays exponentially with an increasing communication distance $L$. However, larger fidelity with a lower error rate always yields a larger $R$, as shown in Fig.~\ref{Fig2}. For a threshold fidelity of $P_{\rm th}=0.943$, the QSS generation rate is $R=0$, 
indicating that the three communication parties can share a private key through channels with fidelities larger than $P_{\rm th}$.

In the previous discussion, the influence of the dark count of  single-photon detectors was neglected. In fact, two types of dark counts  contribute to the error rate: (I) Two of the four single-photon detectors simultaneously click due to the dark counts, and the BSA result aligns with the correlation described in Tables~\ref{table1} and~\ref{table2}; (II) One detector clicks due to a dark count rather than being triggered by the expected photon. Therefore, the additional error rates introduced by dark counts are
\begin{eqnarray}
	e_{d_1}&=& 4p_d^2(1 - {p _d})^2, \nonumber \\
	e_{d_2}&=& 4\eta_t^2\eta _D{\eta _d}p_d(1 - {p _d})^2(1-\eta_t^2\eta_d),
\end{eqnarray}
where $p_d$ is the dark-count rate of the single-photon detector, assumed to be identical for four detectors for simplicity.
The total error rate  $e_{\rm tot}$, considering both dark counts and bit-flip errors $e$ in transmission channels, is
\begin{equation}
e_{\rm tot} = \frac{e_{d_1} + 0.5e_{d_2}+ eQ}{e_{d_1} + e_{d_2} + Q}.
	\end{equation}

\begin{figure}[t!]
	\includegraphics[width=0.48\textwidth]{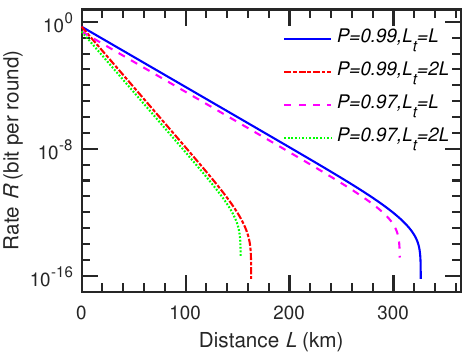}
	\caption{ QSS generation rate $R$ 
of our protocol versus the communication distance 
$L$ with dark counts. We set the dark-count rate to  {$p_d={10^{ - 7}}$}, and the other parameters are consistent with those used in Fig.~\ref{Fig2}. }
	\label{Fig3}
\end{figure}

The modified QSS generation rate of our MDI-QSS protocol can be obtained by substituting $h(e)$ in Eq.~(\ref{Rqss}) with $h(e_{\rm tot})$. The modified QSS generation rate $R$ as a function of the communication distance 
is shown in Fig.~\ref{Fig3}, where we assume a practical dark-count rate of $p_d={10^{- 7}}$, and the other parameters are ${\eta _D} = 86.3\% $, ${\eta _d} = 93\% $,  and  $\alpha  = 0.19~{{{\rm{dB}}} \mathord{\left/{\vphantom {{{\rm{dB}}} {{\rm{km}}}}} \right.\kern-\nulldelimiterspace} {{\rm{km}}}}$~\cite{Li2023High-rate,Marsili2013Detecting,Li2023Breaking}. The communication distance $L$ reaches its limit at approximately  {$163$~km} and  {$153$~km}  for $P=0.99$ and $P=0.97$,  respectively, as shown in Fig.~\ref{Fig3}.

The preceding  analysis assumes that the four quantum channels connecting  Alice and David, mediated by Bob and Charlie, are of  equal length $L$ and each photon transmits a distance of $L_t=2L$. For a modified communication framework, in which nodes Bob and Charlie are closer to node David, the channel losses in these two channels  can thus be neglected, and each photon transmits a distance of $L_t=L$. The dark count rate of $e_{d_2}$ is modified to $e'_{d_2}=4\eta_t\eta _D{\eta _d}p_d(1 - {p_d})^2(1-\eta_t\eta_d)$, while $e_{d_1}$ remains unchanged.
The QSS  generation rate can be calculated accordingly. The maximum transmission distance extends further to approximately $326$~km and $306$~km for $P=0.99$ and $P=0.97$, respectively, as shown in Fig.~\ref{Fig3}. Recently, the distribution of polarization-entangled $1324$-nm telecom photons with fidelity $P=0.99$ has been demonstrated over deployed fibers with 17.4 dB loss~\cite{Craddock2024Automated}. This suggests that the distribution of polarization-entangled $1550$-nm photons with fidelity $P\geq0.97$ over distances up to $300$~km should be feasible in the near future using ultra-low-loss fibers~\cite{Xu2020Secure}.
 
\section{Discussion and summary}

In our work, we use an entangled photon pair as a data bus connecting all communication parties, who randomly perform one of four conjugate unitary operations to participate in the MDI-QSS protocol. The communication distance of our protocol can, in principle, be extended by incorporating quantum memories~\cite{Knaut2024Entanglement,Wang2019Efficient} to eliminate the synchronization required for performing Bell-state analysis based on linear optical elements~\cite{Sheng2010Complete,Ren2013Hyperentanglement,Bhaskar2020Experimental}. Long-distance MDI-QSS is possible when additional entangled channels are preset and incorporated into our MDI-QSS via quantum repeaters~\cite{Azuma2023Quantum,Munro2015Inside,Dias2022Distributing,Jiang2024Efficient,Su2024Heralded}. The untrusted ancillary node in our protocol always performs measurements using the BSA rather than the more complex GSA, as shown in the~\ref{appB}. This leads to an upper bound efficiency of $1/2$, which is independent of the number of communication parties and thus is significantly important for intracity multiparty quantum communications. In contrast, the upper bound efficiency of MDI-QSS protocols based on the GSA decreases by $1/2$ when an additional party is introduced~\cite{Fu2015Long-Distance,Gao2020Deterministic, Das2021Universal, Wei2022Sender-controlled,Ju2022Measurement-device-independent,Li2023Breaking, zhang2024memory}. The number of photons interfering simultaneously at a GSA equals the number of parties~\cite{Fu2015Long-Distance,Gao2020Deterministic, Das2021Universal, Wei2022Sender-controlled,Ju2022Measurement-device-independent,Li2023Breaking, zhang2024memory}. This efficiency decrease corresponds to a loss of approximately $3$ dB, which, in practice, corresponds to transmitting a photon over a fiber of length $15.8$ km with an attenuation coefficient of $0.19$ dB/km. Meanwhile, the communication parties can perform the MDI-QSS protocol without the requirement of single-photon generation~\cite{Hochrainer2022Quantum,Tang2021Towards,Azuma2024Heralded} and thus the interference of single photons emerging from independent sources~\cite{Illiano2024Quantum,Chen2024Entanglement,Zhao2025Direct}.

In summary, we have presented an MDI-QSS protocol using  a two-photon BSA. The parties encode private information by performing conjugate operations on an entangled photon pair instead of preparing single photons in conjugate bases. Our MDI-QSS protocol can therefore be flexibly extended to incorporate more parties without decreasing the efficiency by postselecting two-photon Bell states with efficiency $1/2$, independent of the party number. Furthermore, we have presented the upper bound of the mutual information between the communication parties and the eavesdropper and shown the analytical QSS generation rate among three parties. We believe that the MDI-QSS protocol can contribute significantly to the development of multiparty quantum communication technologies.

\section*{Acknowledgments}
This work was supported by the National Natural Science Foundation of China (Grants No. 11904171 and
No. 62221004).

%\appendices   %仅一个附录时用appendix，否则\appendices
%\section*{APPENDICES} 
%\section*{APPENDIX} 

\setcounter{table}{0}   %从0开始编号，显示出来表会A1开始编号
\setcounter{figure}{0}
\setcounter{section}{0}
\setcounter{equation}{0}
%定义编号格式，在数字序号前加字符“A"
\renewcommand{\thetable}{A\arabic{table}}
\renewcommand{\thefigure}{A\arabic{figure}}
\renewcommand{\thesection}{APPENDIX \Alph{section}}
\renewcommand{\theequation}{A\arabic{equation}}

\section{{Derivation of the mutual information $I(A:E)$}} \label{appA}

%\section {Appendix}
%\appendix   %仅一个附录时用appendix，否则\appendices
%\setcounter{table}{0}   %从0开始编号，显示出来表会A1开始编号
%\setcounter{figure}{0}
%\setcounter{section}{0}
%\setcounter{equation}{0}
%%定义编号格式，在数字序号前加字符“A"
%\renewcommand{\thetable}{A\arabic{table}}
%\renewcommand{\thefigure}{A\arabic{figure}}
%\renewcommand{\thesection}{A\arabic{section}}
%\renewcommand{\theequation}{A\arabic{equation}}
The upper bound of mutual information between Alice and Eve can be obtained by using the Hovelo theorem~\cite{holevo1973bounds} and then can be described as
\begin{equation}
	\begin{aligned}
		I(A:E)& \le \mathop {\max }\limits_{\{ U\} } \{ S({\rho ^{ACE}}) - \frac{1}{4}S(\rho _{00}^{AE})- \frac{1}{4}S(\rho _{01}^{AE}) 	\\
		&~~~- \frac{1}{4}S(\rho _{10}^{AE}) - \frac{1}{4}S(\rho _{11}^{AE})\},
	\end{aligned}
\end{equation}
where $\rho _{ab}^{AE} = {U_{ab}}{U_C}(\rho  \otimes \left| \varepsilon  \right\rangle \left\langle \varepsilon  \right|)U_C^\dag U_{ab}^\dag$ with different subscripts are determined by the unitary operations  $U_{ab}$ of Charlie, and $ S({\rho ^{ACE}})$ is the combined state of Alice, Charlie, and Eve, shown in Eq.~(\ref{rhoACE}). We have $S(\rho _{ab}^{AE}) = S(\rho ) = 1$~{\cite{qi2019implementation}}, 
since $\rho _{ab}^{AE}$ and ${\rho ^{AE}}$ can be converted to  $\rho  \otimes \left| \varepsilon  \right\rangle \left\langle \varepsilon  \right|$  by unitary transformations and the von Neumann entropy of $\rho$, shown in Eq.~(\ref{eq4}), is $S(\rho ) = 1$. Therefore, the  mutual information between Alice and Eve can be simplified to 
\begin{equation}
	I(A:E) \le \mathop {\max }\limits_{\{ U\} } \{ S({\rho ^{ACE}})\}  - 1,
\end{equation}
which is limited by the density matrix $\rho ^{ACE}$.

To obtain $\rho ^{ACE}$, we consider  state evolutions of the photons sent to Charlie and express four Bell states, shown in Eq.~(\ref{eq1}), as follows:
\begin{subequations}
	\begin{align}
	\left| {{\psi ^ + }} \right\rangle  &= \frac{1}{{\sqrt 2 }}({\left|  +  \right\rangle _B}{\left|  +  \right\rangle _C} - {\left|  -  \right\rangle _B}{\left|  -  \right\rangle _C}),\\
	\left| {{\psi ^ - }} \right\rangle  &= \frac{1}{{\sqrt 2 }}( - {\left|  +  \right\rangle _B}{\left|  -  \right\rangle _C} + {\left|  -  \right\rangle _B}{\left|  +  \right\rangle _C}),\\
	\left| {{\varphi ^ + }} \right\rangle  &= \frac{i}{{\sqrt 2 }}({\left|  +  \right\rangle _B}{\left| { - i} \right\rangle _C} - {\left|  -  \right\rangle _B}{\left| { + i} \right\rangle _C}),\\
	\left| {{\varphi ^ - }} \right\rangle  &=  - \frac{i}{{\sqrt 2 }}({\left|  +  \right\rangle _B}{\left| { + i} \right\rangle _C} - {\left|  -  \right\rangle _B}{\left| { - i} \right\rangle _C}),
	\end{align}
\end{subequations}
where  two basis states in the $X$-basis and $Y$-basis are $\left|  \pm  \right\rangle=(\left|  H  \right\rangle\pm\left|  V  \right\rangle)/\sqrt{2}$ and $\left|  \pm i  \right\rangle=(\left|  H  \right\rangle\pm i\left|  V  \right\rangle)/\sqrt{2}$, respectively. Meanwhile, Eve's unitary operation $U_C$ transforms four different states as follows:
\begin{subequations}
	\begin{align}
	{U_C}\left|  +  \right\rangle \left| \varepsilon  \right\rangle  &= \left|  +  \right\rangle \left| {{\varepsilon _{ +  + }}} \right\rangle  + \left|  -  \right\rangle \left| {{\varepsilon _{ +  - }}} \right\rangle ,  \\
	{U_C}\left|  -  \right\rangle \left| \varepsilon  \right\rangle  &= \left|  +  \right\rangle \left| {{\varepsilon _{ -  + }}} \right\rangle  + \left|  -  \right\rangle \left| {{\varepsilon _{ -  - }}} \right\rangle , \\
	{U_C}\left| { + i} \right\rangle \left| \varepsilon  \right\rangle  &= \left| { + i} \right\rangle \left| {{\varepsilon _{ + y - y}}} \right\rangle  + \left| { - i} \right\rangle \left| {{\varepsilon _{ + y + y}}} \right\rangle ,  \\
	{U_C}\left| { - i} \right\rangle \left| \varepsilon  \right\rangle  &= \left| { + i} \right\rangle \left| {{\varepsilon _{ - y - y}}} \right\rangle  + \left| { - i} \right\rangle \left| {{\varepsilon _{ - y + y}}} \right\rangle. 
	\end{align}
\end{subequations}
For two quantum states in the $X$-basis, due to the orthogonality and normalization of unitary transformations, we have the following equations:
\begin{subequations}
	\begin{align}
			\left\langle {{{\varepsilon _{ +  + }}}}
			\mathrel{\left | {\vphantom {{{\varepsilon _{ +  + }}} {{\varepsilon _{ +  + }}}}}
				\right. \kern-\nulldelimiterspace}
			{{{\varepsilon _{ +  + }}}} \right\rangle  + \left\langle {{{\varepsilon _{ +  - }}}}
			\mathrel{\left | {\vphantom {{{\varepsilon _{ +  - }}} {{\varepsilon _{ +  - }}}}}
				\right. \kern-\nulldelimiterspace}
			{{{\varepsilon _{ +  - }}}} \right\rangle  &= 1,\\
			\left\langle {{{\varepsilon _{ -  - }}}}
			\mathrel{\left | {\vphantom {{{\varepsilon _{ -  - }}} {{\varepsilon _{ -  - }}}}}
				\right. \kern-\nulldelimiterspace}
			{{{\varepsilon _{ -  - }}}} \right\rangle  + \left\langle {{{\varepsilon _{- +  }}}}
			\mathrel{\left | {\vphantom {{{\varepsilon _{ - +  }}} {{\varepsilon _{ - +  }}}}}
				\right. \kern-\nulldelimiterspace}
			{{{\varepsilon _{- +  }}}} \right\rangle  &= 1,\\
			\left\langle {{{\varepsilon _{ +  + }}}}
			\mathrel{\left | {\vphantom {{{\varepsilon _{ +  + }}} {{\varepsilon _{- +  }}}}}
				\right. \kern-\nulldelimiterspace}
			{{{\varepsilon _{- +  }}}} \right\rangle  + \left\langle {{{\varepsilon _{ + -  }}}}
			\mathrel{\left | {\vphantom {{{\varepsilon _{ + -  }}} {{\varepsilon _{ -  - }}}}}
				\right. \kern-\nulldelimiterspace}
			{{{\varepsilon _{ -  - }}}} \right\rangle  &= 0. \label{xorthogonal}
	\end{align}
\end{subequations}
Similarly, for two quantum states in the $Y$-basis, we have
\begin{subequations}
	\begin{align}
	\left\langle {{{\varepsilon _{ + y - y}}}}
	\mathrel{\left | {\vphantom {{{\varepsilon _{ + y - y}}} {{\varepsilon _{ + y - y}}}}}
		\right. \kern-\nulldelimiterspace}
	{{{\varepsilon _{ + y - y}}}} \right\rangle  + \left\langle {{{\varepsilon _{ + y + y}}}}
	\mathrel{\left | {\vphantom {{{\varepsilon _{ + y + y}}} {{\varepsilon _{ + y + y}}}}}
		\right. \kern-\nulldelimiterspace}
	{{{\varepsilon _{ + y + y}}}} \right\rangle  = 1,\\
	\left\langle {{{\varepsilon _{ - y + y}}}}
	\mathrel{\left | {\vphantom {{{\varepsilon _{ - y + y}}} {{\varepsilon _{ - y + y}}}}}
		\right. \kern-\nulldelimiterspace}
	{{{\varepsilon _{ - y + y}}}} \right\rangle  + \left\langle {{{\varepsilon _{ - y - y}}}}
	\mathrel{\left | {\vphantom {{{\varepsilon _{ - y - y}}} {{\varepsilon _{ - y - y}}}}}
		\right. \kern-\nulldelimiterspace}
	{{{\varepsilon _{ - y - y}}}} \right\rangle  = 1,\\
	\left\langle {{{\varepsilon _{ + y - y}}}}
	\mathrel{\left | {\vphantom {{{\varepsilon _{ + y - y}}} {{\varepsilon _{ - y - y}}}}}
		\right. \kern-\nulldelimiterspace}
	{{{\varepsilon _{- y - y}}}} \right\rangle  + \left\langle {{{\varepsilon _{  + y + y}}}}
	\mathrel{\left | {\vphantom {{{\varepsilon _{ + y + y}}} {{\varepsilon _{ - y + y}}}}}
		\right. \kern-\nulldelimiterspace}
	{{{\varepsilon _{ - y + y}}}} \right\rangle  = 0.		
	\end{align}
\end{subequations}

The density matrix $\rho ^{ACE}$ can thus be represented as a mixed state consisting of eight pure quantum states, when Charlie randomly applies four encoding unitary operations $U_{ab}$ on one photon of the entangled photon pair that are in state $\left| {{\varphi ^ + }} \right\rangle$~($\left| {{\psi ^ + }} \right\rangle$) or $ \left| {{\varphi ^ - }} \right\rangle$~($\left| {{\psi ^ - }} \right\rangle$) with an equal probability, and can be described as 
\begin{equation}
{\rho ^{ACE}} = \sum\limits_{i = 0}^7 {{p_i}} \left| {{\phi _i}} \right\rangle \left\langle {{\phi _i}} \right|, \label{rhoACEapp}
\end{equation}
where the state $\left| {{\phi _i}} \right\rangle$ for $i = 2(2a+b)$ can be represented as
\begin{equation}
\left| {{\phi _i}} \right\rangle = \frac{1}{2}{U_{ab}}{U_C}(\left|  +  \right\rangle {\left|  +  \right\rangle _C} - \left|  -  \right\rangle {\left|  -  \right\rangle _C})\left| \varepsilon  \right\rangle,
\end{equation}
while the state $\left| {{\phi _i}} \right\rangle$ for $i = 2(2a+b)+1$  can be represented as
\begin{equation}
\left| {{\phi _i}} \right\rangle = \frac{1}{2}{U_{ab}}{U_C}( - \left|  +  \right\rangle {\left|  -  \right\rangle _C} + \left|  -  \right\rangle {\left|  +  \right\rangle _C})\left| \varepsilon  \right\rangle.
\end{equation}
Specifically, these eight quantum states can be detailed as follows:
\begin{subequations}
	\begin{align}
	\left| {{\phi _0}} \right\rangle  &= \frac{1}{2}[\left|  +  \right\rangle ({\left|  +  \right\rangle _C}\left| {{\varepsilon _{ +  + }}} \right\rangle  + {\left|  -  \right\rangle _C}\left| {{\varepsilon _{ +  - }}} \right\rangle )\nonumber\\
	&~~~- \left|  -  \right\rangle ({\left|  +  \right\rangle _C}\left| {{\varepsilon _{ -  + }}} \right\rangle  + {\left|  -  \right\rangle _C}\left| {{\varepsilon _{ -  - }}} \right\rangle )],\\
	\left| {{\phi _1}} \right\rangle  &=  - \frac{1}{2}[\left|  +  \right\rangle ({\left|  +  \right\rangle _C}\left| {{\varepsilon _{ -  + }}} \right\rangle  + {\left|  -  \right\rangle _C}\left| {{\varepsilon _{ -  - }}} \right\rangle )\nonumber\\
	&~~~+ \left|  -  \right\rangle ({\left|  +  \right\rangle _C}\left| {{\varepsilon _{ +  + }}} \right\rangle  + {\left|  -  \right\rangle _C}\left| {{\varepsilon _{ +  - }}} \right\rangle )],\\
	\left| {{\phi _2}} \right\rangle  &= \frac{1}{2}[\left|  +  \right\rangle ({\left|  -  \right\rangle _C}\left| {{\varepsilon _{ +  + }}} \right\rangle  + {\left|  +  \right\rangle _C}\left| {{\varepsilon _{ +  - }}} \right\rangle )\nonumber\\
	&~~~- \left|  -  \right\rangle ({\left|  -  \right\rangle _C}\left| {{\varepsilon _{ -  + }}} \right\rangle  + {\left|  +  \right\rangle _C}\left| {{\varepsilon _{ -  - }}} \right\rangle )],\\
	\left| {{\phi _3}} \right\rangle  &=  - \frac{1}{2}[\left|  +  \right\rangle ({\left|  -  \right\rangle _C}\left| {{\varepsilon _{ -  + }}} \right\rangle  + {\left|  +  \right\rangle _C}\left| {{\varepsilon _{ -  - }}} \right\rangle )\nonumber\\
	&~~~+ \left|  -  \right\rangle ({\left|  -  \right\rangle _C}\left| {{\varepsilon _{ +  + }}} \right\rangle  + {\left|  +  \right\rangle _C}\left| {{\varepsilon _{ +  - }}} \right\rangle )],\\
	\left| {{\phi _4}} \right\rangle  &= \frac{1}{2}[\left|  +  \right\rangle ({\left|  -  \right\rangle _C}\left| {{\varepsilon _{ - y - y}}} \right\rangle  + {\left|  +  \right\rangle _C}\left| {{\varepsilon _{ - y + y}}} \right\rangle )\nonumber\\
	&~~~- \left|  -  \right\rangle ({\left|  -  \right\rangle _C}\left| {{\varepsilon _{ + y - y}}} \right\rangle  + {\left|  +  \right\rangle _C}\left| {{\varepsilon _{ + y + y}}} \right\rangle )],\\
	\left| {{\phi _5}} \right\rangle  &=  - \frac{1}{2}[\left|  +  \right\rangle ({\left|  -  \right\rangle _C}\left| {{\varepsilon _{ + y - y}}} \right\rangle  + {\left|  +  \right\rangle _C}\left| {{\varepsilon _{ + y + y}}} \right\rangle )\nonumber\\
&~~~	+ \left|  -  \right\rangle ({\left|  -  \right\rangle _C}\left| {{\varepsilon _{ - y - y}}} \right\rangle  + {\left|  +  \right\rangle _C}\left| {{\varepsilon _{ - y + y}}} \right\rangle )],\\
	\left| {{\phi _6}} \right\rangle  &= \frac{1}{2}[\left|  +  \right\rangle ({\left|  +  \right\rangle _C}\left| {{\varepsilon _{ - y - y}}} \right\rangle  + {\left|  -  \right\rangle _C}\left| {{\varepsilon _{ - y + y}}} \right\rangle )\nonumber\\
	&~~~- \left|  -  \right\rangle ({\left|  +  \right\rangle _C}\left| {{\varepsilon _{ + y - y}}} \right\rangle  + {\left|  -  \right\rangle _C}\left| {{\varepsilon _{ + y + y}}} \right\rangle )],\\
	\left| {{\phi _7}} \right\rangle  &=  - \frac{1}{2}[\left|  +  \right\rangle ({\left|  +  \right\rangle _C}\left| {{\varepsilon _{ + y - y}}} \right\rangle  + {\left|  -  \right\rangle _C}\left| {{\varepsilon _{ + y + y}}} \right\rangle )\nonumber\\
&~~~	+ \left|  -  \right\rangle ({\left|  +  \right\rangle _C}\left| {{\varepsilon _{ - y - y}}} \right\rangle  + {\left|  -  \right\rangle _C}\left| {{\varepsilon _{ - y + y}}} \right\rangle )].
\end{align}
\end{subequations}

The corresponding Gram matrix of ${\rho ^{ACE}}$, shown in Eq.~(\ref{rhoACEapp}), with $G_{ij}=\sqrt{p_ip_j}\langle \phi_i|\phi_j\rangle${\cite{PhysRevA.62.012301}}, can be explicitly represented as:
\begin{widetext} 
\begin{equation}
G = \frac{1}{{16}}\left( {\begin{array}{*{20}{c}}
		2&0&{2\alpha }&{ - 2\beta }&{1 + \beta  - i\alpha }&{ - \alpha  + i - i\beta }&{\alpha  + i - i\beta }&{ - 1 - \beta  - i\alpha }\\
		0&2&{ - 2\beta }&{2\alpha }&{ - \alpha  + i - i\beta }&{1 + \beta  - i\alpha }&{ - 1 - \beta  - i\alpha }&{\alpha  + i - i\beta }\\
		{2\alpha }&{ - 2\beta }&2&0&{\alpha  + i - i\beta }&{ - 1 - \beta  - i\alpha }&{1 + \beta  - i\alpha }&{ - \alpha  + i - i\beta }\\
		{ - 2\beta }&{2\alpha }&0&2&{ - 1 - \beta  - i\alpha }&{\alpha  + i - i\beta }&{ - \alpha  + i - i\beta }&{1 + \beta  - i\alpha }\\
		{1 + \beta  + i\alpha }&{ - \alpha  - i + i\beta }&{\alpha  - i + i\beta }&{ - 1 - \beta  + i\alpha }&2&0&{2\alpha }&{ - 2\beta }\\
		{ - \alpha  - i + i\beta }&{1 + \beta  + i\alpha }&{ - 1 - \beta  + i\alpha }&{\alpha  - i + i\beta }&0&2&{ - 2\beta }&{2\alpha }\\
		{\alpha  - i + i\beta }&{ - 1 - \beta  + i\alpha }&{1 + \beta  + i\alpha }&{ - \alpha  - i + i\beta }&{2\alpha }&{ - 2\beta }&2&0\\
		{ - 1 - \beta  + i\alpha }&{\alpha  - i + i\beta }&{ - \alpha  - i + i\beta }&{1 + \beta  + i\alpha }&{ - 2\beta }&{2\alpha }&0&2
\end{array}} \right),
\end{equation}
\end{widetext}
where the auxiliary parameters are $\alpha = {\mathop{\rm Re}\nolimits} (\left\langle {{{\varepsilon _{ +  + }}}}
\mathrel{\left | {\vphantom {{{\varepsilon _{ +  + }}} {{\varepsilon _{ +  - }}}}}
	\right. \kern-\nulldelimiterspace}
{{{\varepsilon _{ +  - }}}} \right\rangle  + \left\langle {{{\varepsilon _{ -  + }}}}
\mathrel{\left | {\vphantom {{{\varepsilon _{ -  + }}} {{\varepsilon _{ -  - }}}}}
	\right. \kern-\nulldelimiterspace}
{{{\varepsilon _{ -  - }}}} \right\rangle )=0$ and $\beta={\mathop{\rm Re}\nolimits} (\left\langle {{{\varepsilon _{ +  + }}}}
\mathrel{\left | {\vphantom {{{\varepsilon _{ +  + }}} {{\varepsilon _{ -  - }}}}}
	\right. \kern-\nulldelimiterspace}
{{{\varepsilon _{ -  - }}}} \right\rangle  + \left\langle {{{\varepsilon _{ +  - }}}}
\mathrel{\left | {\vphantom {{{\varepsilon _{ +  - }}} {{\varepsilon _{ -  + }}}}}
	\right. \kern-\nulldelimiterspace}
{{{\varepsilon _{ -  + }}}} \right\rangle )$. The eight eigenvalues of Gram matrix $G$ can be directly obtained by solving the characteristic
function det$|G-\lambda_iI|=0$ with $I$ being the $8\times8$ identity matrix, and can be specified as follows:
\begin{subequations}
	\begin{align}
{\lambda _1} &= {\lambda _2} = {\lambda _3} ={\lambda _4} = 0,\\
{\lambda _5} &= {\lambda _6} =\frac{1}{4}(1 - \beta ),\\
{\lambda _7} &={\lambda _8}= \frac{1}{4}(1 + \beta ).
\end{align}
\end{subequations}

The von Neumann entropy ${S(\rho ^{ACE})}=\sum_{i=1}^{8}\lambda_i\rm{log}_2\lambda_i$ is thus a function of $\beta$ and is determined by the quantum-bit-error rates $e_x$ and $e_y$ for $X$-basis and $Y$-basis states:
\begin{subequations}
	\begin{align}
	{e_x} &= \left\langle {{{\varepsilon _{ +  - }}}}
	\mathrel{\left | {\vphantom {{{\varepsilon _{ +  - }}} {{\varepsilon _{ +  - }}}}}
		\right. \kern-\nulldelimiterspace}
	{{{\varepsilon _{ +  - }}}} \right\rangle
  = \left\langle {{{\varepsilon _{ -  + }}}}
	\mathrel{\left | {\vphantom {{{\varepsilon _{ -  + }}} {{\varepsilon _{ - + }}}}}
		\right. \kern-\nulldelimiterspace}
	{{{\varepsilon _{ - + }}}} \right\rangle ,\\
	{e_y} &= \left\langle {{{\varepsilon _{ + y + y}}}}
	\mathrel{\left | {\vphantom {{{\varepsilon _{ + y + y}}} {{\varepsilon _{ + y + y}}}}}
		\right. \kern-\nulldelimiterspace}
	{{{\varepsilon _{ + y + y}}}} \right\rangle  = \left\langle {{{\varepsilon _{ - y - y}}}}
	\mathrel{\left | {\vphantom {{{\varepsilon _{ - y - y}}} {{\varepsilon _{ - y - y}}}}}
		\right. \kern-\nulldelimiterspace}
	{{{\varepsilon _{ - y - y}}}} \right\rangle \\
	&= \frac{1}{2}[{{1 + {\mathop{\rm Re}\nolimits} (\left\langle {{{\varepsilon _{ +  - }}}}
			\mathrel{\left | {\vphantom {{{\varepsilon _{ +  - }}} {{\varepsilon _{ -  + }}}}}
				\right. \kern-\nulldelimiterspace}
			{{{\varepsilon _{ -  + }}}} \right\rangle  - \left\langle {{{\varepsilon _{ +  + }}}}
			\mathrel{\left | {\vphantom {{{\varepsilon _{ +  + }}} {{\varepsilon _{ -  - }}}}}
				\right. \kern-\nulldelimiterspace}
			{{{\varepsilon _{ -  - }}}} \right\rangle )}} ] \\
	&= \frac{1}{2}[{{1 - \beta  + 2{\mathop{\rm Re}\nolimits} \left\langle {{{\varepsilon _{ +  - }}}}
			\mathrel{\left | {\vphantom {{{\varepsilon _{ +  - }}} {{\varepsilon _{ -  + }}}}}
				\right. \kern-\nulldelimiterspace}
			{{{\varepsilon _{ -  + }}}} \right\rangle }}],
	\end{align}
\end{subequations}
According to the Cauchy–Schwarz inequality, we have ${\left| {\left\langle {{{\varepsilon _{ +  - }}}}
		\mathrel{\left | {\vphantom {{{\varepsilon _{ +  - }}} {{\varepsilon _{ -  + }}}}}
			\right. \kern-\nulldelimiterspace}
		{{{\varepsilon _{ -  + }}}} \right\rangle } \right|^2} \le \left| {\left\langle {{{\varepsilon _{ +  - }}}}
	\mathrel{\left | {\vphantom {{{\varepsilon _{ +  - }}} {{\varepsilon _{ +  - }}}}}
		\right. \kern-\nulldelimiterspace}
	{{{\varepsilon _{ +  - }}}} \right\rangle } \right| \left| {\left\langle {{{\varepsilon _{ -  + }}}}
	\mathrel{\left | {\vphantom {{{\varepsilon _{ -  + }}} {{\varepsilon _{ -  + }}}}}
		\right. \kern-\nulldelimiterspace}
	{{{\varepsilon _{ -  + }}}} \right\rangle } \right|$, 
and then $- {e_x}\le {\rm Re}\langle \varepsilon_{ + - }| \varepsilon_{ -  + }\rangle\le e_x$, leading to the bounds for $\beta$ with  $1 - 2{e_x} - 2{e_y} \le \beta  \le 1 + 2{e_x} - 2{e_y}$.
Under the condition that guarantees ${e_x} + {e_y} \le {1 \mathord{\left/
		{\vphantom {1 2}} \right.
		\kern-\nulldelimiterspace} 2}$, the maximum von Neumann entropy  $S{({\rho ^{ACE}})_{\max }}$  is achieved for $\beta = 1 - 2{e_x} - 2{e_y}$ with $S{({\rho ^{ACE}})_{\max }} = 1 + h({e_x} + {e_y})$. Therefore, the mutual information between Alice and Eve is 
\begin{eqnarray}
I(A:E) = h({e_x} + {e_y}),
\end{eqnarray}
which decreases the QSS generation rate $R$, as shown in Eq.~(\ref{keyRate}).

\section{{Equivalence to multiphoton MDI-QSS protocols}} \label{appB}
\setcounter{equation}{0}
\renewcommand{\theequation}{B\arabic{equation}}
Our three-party MDI-QSS protocol with the BSA can be linked to a MDI-QSS protocol~\cite{Fu2015Long-Distance,Gao2020Deterministic, Das2021Universal, Wei2022Sender-controlled,Ju2022Measurement-device-independent,Li2023Breaking, zhang2024memory} that  postselects a virtual three-photon GHZ state by the purification of Alice's state $\rho$ with an ancillary system~\cite{Xu2020Secure}. Specifically, Alice prepares a virtual three-photon GHZ state $\left| {{\Psi_+}} \right\rangle = \frac{1}{{\sqrt 2 }}(\left| {HHV} \right\rangle  + \left| {VVH} \right\rangle )$, which can be redescribed as
	\begin{align}
	\left| \Psi_+\right\rangle  &= \frac{1}{{\sqrt 2 }}(\left|+\right\rangle\left|\psi^+ \right\rangle+  \left|-\right\rangle\left| \psi^-\right\rangle )\nonumber\\
	&= \frac{1}{{\sqrt 2 }}(\left|+i\right\rangle\left|\varphi^- \right\rangle+  \left|-i\right\rangle\left| \varphi^+\right\rangle ),
	\label{eqGHZ2}
\end{align}
which leads to the mixed state $\rho$ after Alice has measured the ancillary photon in the $X$ ($Y$) basis. The other two photons will be projected onto the Bell state $\left|\psi^+ \right\rangle$~($\left|\varphi^+ \right\rangle$) or $\left|\psi^- \right\rangle$~($\left|\varphi^- \right\rangle$) when the result of the photon measurement is $\left|+\right\rangle$~($\left|-i\right\rangle$) or  $\left|-\right\rangle$~($\left|+i\right\rangle$), respectively.

%\\ \indent 
The measurement on the ancillary photon and the BSA measurement on the two photons, after rotations by Bob and Charlie to incorporate their joint private classical information, can be reordered without modifying the statistics of the measurement results, as shown in the original MDI-QKD protocols~\cite{lo2012measurement,Braunstein2012Side-channel}. 
In our three-party MDI-QSS protocol, the BSA is designed to postselect the Bell states $\left|\psi^+ \right\rangle$ and $\left|\psi^- \right\rangle$. This corresponds to postselecting the virtual three-photon GHZ states $\left| {{\Psi_{\pm}}} \right\rangle = \frac{1}{{\sqrt 2 }}(\left| {HHV} \right\rangle  \pm \left| {VVH} \right\rangle )$ when Alice performs the $X$-basis measurement, or the GHZ states $\left| {{\Psi_{\pm i}}} \right\rangle = \frac{1}{{\sqrt 2 }}(\left| {HHV} \right\rangle  \pm i\left| {VVH} \right\rangle )$ when Alice performs the $Y$-basis measurement. Note that we have implicitly incorporated the restraint condition for the parties' public keys, which leads to the desired correlations between their private keys when the BSA postselects the Bell states $\left|\psi^+ \right\rangle$ or $\left|\psi^- \right\rangle$.

For the four-party MDI-QSS protocol described in Sec.~\ref{secIIB}, an additional communication party, Daniel, is introduced compared to the three-party MDI-QSS protocol. In principle, Daniel's randomly chosen unitary operations, which encode his private classical information, evolve the virtual three-photon GHZ state $\left| \Psi_{+}\right\rangle$ into a maximally mixed state:
\begin{eqnarray}
    \rho' &= &\frac{1}{2} \left( \left| \Psi_{+} \right\rangle \left\langle \Psi_{+} \right| + \left| \Psi_{-} \right\rangle \left\langle \Psi_{-} \right| \right) \nonumber\\
    &= &\frac{1}{2} \left( \left| \Psi_{+i} \right\rangle \left\langle \Psi_{+i} \right| + \left| \Psi_{-i} \right\rangle \left\langle \Psi_{-i} \right| \right),
\end{eqnarray}
which can be purified by incorporating an ancillary photon~\cite{Xu2020Secure}. This leads to a virtual four-photon GHZ state $\left| \Psi'_{+} \right\rangle = \frac{1}{\sqrt{2}} \left( \left| HHHV \right\rangle + \left| VVVH \right\rangle \right)$.
We can then process Daniel's virtual photon in the same way Alice processes hers. After a BSA measurement on the two photons in the untrusted node David yields a successful result $|\psi^\pm\rangle$, the four communication parties postselect virtual four-photon GHZ states 
$\left| \Psi'_{\pm} \right\rangle = \frac{1}{\sqrt{2}} \left( \left| HHHV \right\rangle \pm \left| VVVH \right\rangle \right)$,
or $ \left| \Psi'_{\pm i} \right\rangle = \frac{1}{\sqrt{2}} \left( \left| HHHV \right\rangle \pm i\left| VVVH \right\rangle \right)$,
when the public keys of Alice and Daniel satisfy the conditions $a_A \oplus a_D = 0$ or $a_A \oplus a_D = 1$, respectively.

%\bibliographystyle{apsrev4-2}
%\bibliography{refer-QSS}


\end{document}